\documentclass[a4paper,11pt]{article}

\usepackage[a4paper,margin=2.4cm]{geometry}
\usepackage{graphicx}
\usepackage{booktabs}
\usepackage{array}
\usepackage{multirow}
\usepackage{subcaption}
\usepackage[table]{xcolor}
\usepackage{amsmath}
\usepackage{enumitem}
\usepackage{xurl}
\usepackage{hyperref}
\usepackage{longtable}
\usepackage{float}
\usepackage{makecell}
\usepackage{pifont}
\usepackage[T1]{fontenc}
\usepackage{newtxtext,newtxmath}

\graphicspath{{./figures/}{./figures/nature_system/}}
\hypersetup{
  colorlinks=true,
  linkcolor=blue!55!black,
  citecolor=blue!55!black,
  urlcolor=blue!55!black
}

\newcommand{\system}{TianJi-Environ}
\newcommand{\wrfchem}{WRF-Chem}
\newcommand{\mda}{MDA8 O$_3$}
\newcommand{\pmfine}{PM$_{2.5}$}
\newcommand{\nox}{NO$_x$}
\newcommand{\voc}{VOC}
\newcommand{\pblh}{PBLH}
\newcommand{\swdown}{SWDOWN}
\newcommand{\cmark}{\ding{51}}
\newcommand{\xmark}{\ding{55}}
\definecolor{rowgray}{RGB}{248,248,248}
\definecolor{tianjiblue}{RGB}{222,237,252}
\newcolumntype{L}[1]{>{\raggedright\arraybackslash}p{#1}}
\title{\textbf{\system: An Autonomous AI Scientist for Atmospheric Environmental Research}}
\author{\makebox[0pt][c]{\parbox{0.98\textwidth}{\centering
Haoluo Zhao$^{1}$, Hongchun Zhang$^{1}$, Nan Li$^{3}$, Jing-Jia Luo$^{1,2}$,\\
Kaikai Zhang$^{1}$, Mengyang Yu$^{1}$, Nan Chen$^{1}$, Tao Song$^{4}$, and Fan Meng$^{1,2,*}$\\[0.75em]
{\small $^{1}$School of Artificial Intelligence, Nanjing University of Information Science and Technology, Nanjing, China}\\
{\small $^{2}$State Key Laboratory of Climate System Prediction and Risk Management (CPRM),}\\
{\small Nanjing University of Information Science and Technology, Nanjing, 210044, China}\\
{\small $^{3}$College of Environmental Science and Engineering, Nanjing University of Information Science and Technology, Nanjing, China}\\
{\small $^{4}$College of Computer Science and Technology, China University of Petroleum, Qingdao, Shandong, China}\\
{\small $^{*}$Corresponding author: Fan Meng. E-mail: \href{mailto:meng@nuist.edu.cn}{meng@nuist.edu.cn}}
}}}
\date{}

\begin{document}
\maketitle

\begin{abstract}
As atmospheric environmental prediction continues to improve, interpretable validation of pollution mechanisms and feedback processes has become a main challenge in atmospheric chemistry. Yet mechanism validation based on complex numerical models still relies heavily on expert knowledge: mechanistic hypotheses must be operationalized into executable experiments, and model outputs must be organized into traceable evidence. We present TianJi-Environ, an auditable AI Scientist for atmospheric-chemistry mechanism validation. TianJi-Environ establishes the first \wrfchem{}-based multi-agent framework that autonomously drives complex atmospheric-chemistry simulations, converting mechanistic hypotheses into executable configurations, testing experiments, and evidence criteria. Using ozone response and particulate-matter feedback as two representative examples, we demonstrate TianJi-Environ's capability for mechanism validation. In a summertime ozone case over the North China Plain, the system detects directionally consistent aerosol--radiation-interaction signals in shortwave radiation and boundary-layer height, but judges the evidence for ozone response to NO$_x$ control to be incomplete. In a wintertime PM$_{2.5}$ case over the Guanzhong Basin, it localizes the unsupported link to insufficient propagation from black-carbon perturbation to particulate response and missing diagnostics of vertical absorptive heating. These results show that TianJi-Environ makes expert-driven mechanism validation explicit, structured, and auditable, offering a reproducible paradigm for multi-agent systems coupled with complex atmospheric-chemistry models.
\end{abstract}


\section{Introduction}

Atmospheric environmental research is increasingly moving beyond concentration prediction toward interpretable validation of pollution mechanisms and control responses. Tropospheric ozone and \pmfine{} remain persistent air-quality challenges, but their responses are governed by coupled photochemistry, emissions, aerosol processes, radiation, and boundary-layer dynamics \cite{seinfeld2016atmospheric,jacob1999atmospheric,sillman1999,duncan2010omi,petaja2016feedback,ding2016blackcarbon}. Prediction accuracy or statistical association alone cannot determine whether a proposed process chain is physically supported. Mechanism-oriented research therefore requires controlled experiments and diagnostic evidence that can distinguish supported signals from incomplete or contradicted explanations.

Atmospheric chemistry models provide a key experimental basis for this task. Models such as \wrfchem{} couple meteorology, gas-phase chemistry, aerosol processes, emissions, and radiative transfer in an online modelling framework \cite{grell2005wrfchem,fast2006wrfchem,zhang2008online,baklanov2014online,skamarock2019wrf}. This capability allows researchers to perturb emissions, enable or disable aerosol feedbacks, and examine how process-level changes propagate into ozone, \pmfine{}, radiation, and boundary-layer diagnostics \cite{yang2022apiarf,wu2019arf,li2020aerosolfeedback,gao2022twoway}. However, the model itself does not complete the scientific reasoning loop. Researchers still need to translate hypotheses into executable configurations, coordinate controlled experiments, select diagnostics, interpret evidence states, and decide whether a mechanism is supported, incomplete, or unsupported. This workflow remains largely expert-driven and difficult to trace or reuse.

Recent AI systems have substantially advanced atmospheric and Earth-system prediction. Data-driven models such as Pangu-Weather, GraphCast, GenCast, Aurora, and AI-GAMFS show that AI can rapidly predict increasingly rich atmospheric states, including weather, air-quality variables, and aerosol-related quantities \cite{bi2023pangu,lam2023graphcast,price2024gencast,bodnar2025aurora,gui2026aigamfs}. Yet prediction is not equivalent to mechanism validation. A mechanism study must ask not only what will happen, but why it happens, which diagnostic evidence supports it, and where the evidential chain breaks.

Large language models and multi-agent systems offer a possible route toward organizing such long-horizon scientific workflows \cite{brown2020language,wei2022chain,achiam2023gpt4,yao2023react,shinn2023reflexion,hong2023metagpt,wu2023autogen,wang2024agents}. Recent AI-scientist and tool-augmented discovery systems have demonstrated capabilities in hypothesis generation, code execution, autonomous experimentation, and scientific reporting \cite{boiko2023autonomous,bran2023chemcrow,lu2024aiscientist,ghafarollahi2024sciagents,wang2023scientific}. However, existing systems mainly operate in domains where experiments are code-level, laboratory-tool-based, or data-analysis-oriented. They have not yet established a stable workflow for connecting atmospheric-chemistry hypotheses with \wrfchem{} execution, diagnostic evidence construction, and mechanism-level interpretation.

Here we introduce \system{}\footnote{We name our AI Scientist TianJi-Environ. In Chinese philosophy, ``Tian'' refers to the sky or heaven, and ``Ji'' denotes mechanisms or profound secrets. Historically, forecasting the weather was considered attempting to decipher ``TianJi''---the unrevealable secrets of nature. Here, our model systematically unravels these atmospheric mechanisms through artificial intelligence. Environ indicates the atmospheric-environmental domain of the system.}, an auditable AI Scientist for atmospheric-chemistry mechanism validation. TianJi-Environ constructs a \wrfchem{}-based multi-agent framework that autonomously drives complex atmospheric-chemistry simulations and transforms mechanistic hypotheses into executable configurations, testing experiments, diagnostics, and evidence criteria. The system records the research process as a traceable hypothesis--experiment--diagnosis--evidence--conclusion workflow. Through two representative problems, ozone response and particulate-matter feedback, we show that TianJi-Environ can identify directionally consistent mechanism signals, detect incomplete evidence chains, and localize unsupported links in hypothesized atmospheric processes.

The main contributions of this work are threefold. First, we formulate atmospheric-chemistry mechanism validation as an auditable AI Scientist task grounded in complex numerical modelling. Second, we develop the first \wrfchem{}-based multi-agent framework for autonomous mechanism-testing experiments and evidence construction. Third, we demonstrate, through ozone and \pmfine{} case studies, that the framework can make expert-driven mechanism validation explicit, structured, and reproducible.

\section{Related Work}

\subsection{Atmospheric Chemistry Models for Mechanism Validation}

Atmospheric chemistry models are essential tools for mechanism-oriented air-quality research. Models such as CMAQ and \wrfchem{} integrate meteorology, chemical transformation, aerosol processes, emissions, deposition, and radiative interactions within a numerical modelling framework \cite{byun2006cmaq,grell2005wrfchem,fast2006wrfchem,zhang2008online,baklanov2014online}. In particular, the online coupling in \wrfchem{} enables feedbacks between meteorology and chemistry to be represented during the same simulation, making it suitable for testing aerosol--radiation interaction, aerosol--photolysis interaction, emission perturbations, and boundary-layer feedbacks \cite{skamarock2019wrf,gao2022twoway}.

Previous studies have used model perturbation and sensitivity experiments to examine ozone formation, aerosol suppression of photolysis, absorbing-aerosol feedback, and \pmfine{} accumulation under stable boundary-layer conditions \cite{yang2022apiarf,wu2019arf,li2020aerosolfeedback,petaja2016feedback,ding2016blackcarbon,wang2018dome,li2022indicators}. These studies show that mechanism validation requires more than pollutant concentration changes: it also depends on diagnostics of radiation, boundary-layer evolution, chemical sensitivity, aerosol composition, and process-level consistency.

Despite their scientific value, model-based mechanism studies remain difficult to automate. Researchers must translate hypotheses into model settings, define controlled contrasts, maintain input consistency, select diagnostics, interpret evidence, and document conclusions. These steps are often distributed across scripts, configuration files, diagnostic notebooks, and informal expert judgement. TianJi-Environ targets this gap by making the model-based mechanism-validation workflow explicit and auditable.

\subsection{LLM Agents and AI Scientists}

Large language models have demonstrated strong capabilities in reasoning, coding, tool use, and task decomposition \cite{brown2020language,wei2022chain,achiam2023gpt4}. Tool-augmented agents extend these capabilities by connecting language reasoning with retrieval, computation, external tools, and iterative feedback \cite{yao2023react,shinn2023reflexion}. Multi-agent frameworks further decompose complex workflows into planning, execution, validation, and reporting, allowing specialized agents to coordinate through shared state and messages \cite{hong2023metagpt,wu2023autogen,wang2024agents}.

Scientific discovery requires more than producing plausible answers. It involves hypothesis formulation, experiment design, execution, evidence assessment, and qualified conclusions under uncertainty. Recent AI-scientist systems and autonomous discovery frameworks have explored this direction in machine learning, chemistry, and materials-related settings \cite{lu2024aiscientist,boiko2023autonomous,bran2023chemcrow,ghafarollahi2024sciagents,wang2023scientific}. However, their experimental environments are typically code-level, laboratory-tool-oriented, or data-analysis-based. Atmospheric-chemistry mechanism validation poses a different challenge: the agent must interact with a complex numerical model, respect domain-specific configuration constraints, and interpret evidence across coupled physical and chemical processes. TianJi-Environ is designed for this domain-grounded form of AI Scientist.

\subsection{AI for Atmospheric and Earth-System Research}

AI has rapidly advanced Earth-system prediction and environmental modelling. Pangu-Weather, GraphCast, and GenCast have demonstrated high-skill global weather prediction \cite{bi2023pangu,lam2023graphcast,price2024gencast}; Aurora extends foundation-model prediction to multiple atmospheric and environmental variables \cite{bodnar2025aurora}; and AI-GAMFS targets coupled aerosol--meteorology forecasting at global scale \cite{gui2026aigamfs}. These systems demonstrate the growing ability of AI to emulate or accelerate atmospheric-state prediction.

Agentic systems for Earth science have also begun to support data analysis, remote-sensing interpretation, and spatiotemporal reasoning \cite{guo2025earthlink,feng2026earthagent}. More broadly, Earth-system AI research has emphasized the need to connect data-driven skill with process understanding \cite{karpatne2017theory,reichstein2019deep}. However, existing agentic systems remain largely focused on prediction, data processing, retrieval, remote-sensing reasoning, or interactive analysis. They do not directly address the problem of autonomously organizing \wrfchem{}-based mechanism-testing experiments and constructing traceable evidence chains. TianJi-Environ complements these efforts by shifting the target from predictive skill to mechanism validation grounded in controlled atmospheric-chemistry simulations.

\begin{table}[H]
\centering
\caption{\textbf{Comparison with representative autonomous scientific discovery and Earth-system agent frameworks.}}
\label{tab:framework_comparison}
\small
\setlength{\tabcolsep}{5pt}
\renewcommand{\arraystretch}{1.18}
\resizebox{\linewidth}{!}{%
\begin{tabular}{lcccccc}
\toprule
\multirow{2}{*}{\textbf{Method}} &
\multicolumn{2}{c}{\textbf{Domains}} &
\multicolumn{4}{c}{\textbf{Capabilities}} \\
\cmidrule(lr){2-3}\cmidrule(lr){4-7}
&
\makecell{\textbf{General}\\\textbf{Discovery}} &
\makecell{\textbf{Atmospheric}\\\textbf{Mechanism}} &
\makecell{\textbf{Literature}\\\textbf{Survey}} &
\makecell{\textbf{Hypothesis}\\\textbf{Design}} &
\makecell{\textbf{WRF-Chem}\\\textbf{Validation}} &
\makecell{\textbf{Scientific}\\\textbf{Report}} \\
\midrule
AI Scientist~\cite{lu2024aiscientist} & \cmark & \xmark & \cmark & \cmark & \xmark & \cmark \\
\rowcolor{rowgray}
ChemCrow~\cite{bran2023chemcrow} & \cmark & \xmark & \cmark & \cmark & \xmark & \xmark \\
Coscientist~\cite{boiko2023autonomous} & \cmark & \xmark & \cmark & \cmark & \xmark & \xmark \\
\rowcolor{rowgray}
EarthLink~\cite{guo2025earthlink} & \cmark & \cmark & \cmark & \cmark & \xmark & \cmark \\
Earth-Agent~\cite{feng2026earthagent} & \cmark & \cmark & \xmark & \xmark & \xmark & \xmark \\
\midrule
\rowcolor{tianjiblue}
\textbf{\system} & \cmark & \cmark & \cmark & \cmark & \cmark & \cmark \\
\bottomrule
\end{tabular}%
}
\end{table}

\section{TianJi-Environ Framework}

\subsection{Mechanism-Validation Workflow}

\begin{figure}[H]
  \centering
  \includegraphics[width=0.98\linewidth]{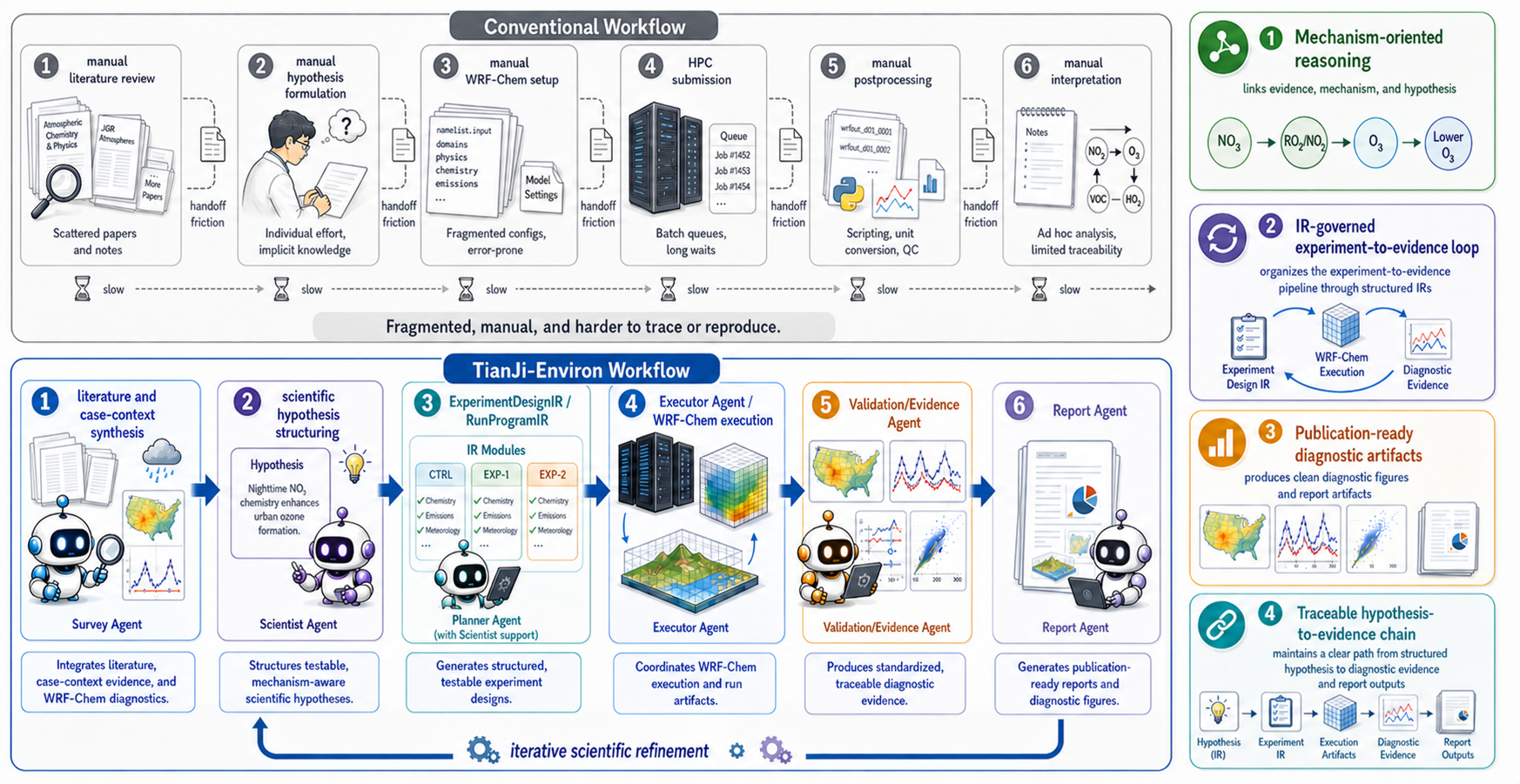}
  \caption{\textbf{Comparison between the traditional atmospheric-chemistry workflow and the autonomous research loop of \system.} The traditional workflow relies on manual coordination among literature reading, hypothesis construction, \wrfchem{} experiment preparation, model execution, post-processing, and interpretation. \system{} organizes these steps as a research loop spanning literature synthesis, hypothesis organization, experiment design, execution coordination, evidence evaluation, and report expression. Validation results and evidence gaps can also be fed back as targeted survey questions for subsequent hypothesis refinement and experiment design.}
  \label{fig:workflow_comparison}
\end{figure}

\system{} is designed to reduce the separation between hypotheses, experiments, and evidence in atmospheric-chemistry mechanism research. Traditional workflows can produce high-quality judgements, but literature evidence, experimental intent, model assumptions, diagnostic figures, and final interpretation are often distributed across different media. When a result deviates from the hypothesis, researchers must reconstruct which assumptions shaped the experiment, which diagnostics are missing, and which conclusions depend on expert judgement. \system{} organizes these stages into a research loop advanced by an AI scientist and reviewed by human researchers, making the provenance of experimental intent and evidential judgement explicit (Fig.~\ref{fig:workflow_comparison}).

This loop relies on a continuously updated research state rather than a one-shot question-answer exchange. Literature survey, hypothesis organization, experiment design, result retrieval, and evidence evaluation are placed within the same research trajectory, preserving the scientific origin, purpose, and downstream dependence of each step. Autonomy in this setting therefore does not mean that the language model freely produces conclusions. It means that research actions can be sequenced according to the current question, available evidence, and remaining uncertainty.

Feedback in this loop does not only occur at the reporting stage. When a validation run exposes missing diagnostics, ineffective perturbations, or a mismatch between the target response and the expected mechanism, these evidence gaps can be translated into targeted literature-survey questions for the next round. For example, subsequent survey can focus on BC vertical distributions, absorption-heating profiles, JNO$_2$, AOD550, or other diagnostics needed to discriminate the mechanism. This result-driven resurvey does not rewrite the current judgement. It carries the uncertainty exposed by validation back into literature evidence, hypothesis refinement, and subsequent experiment design.

\subsection{Multi-Agent Coordination}

\begin{figure}[H]
  \centering
  \includegraphics[width=0.98\linewidth]{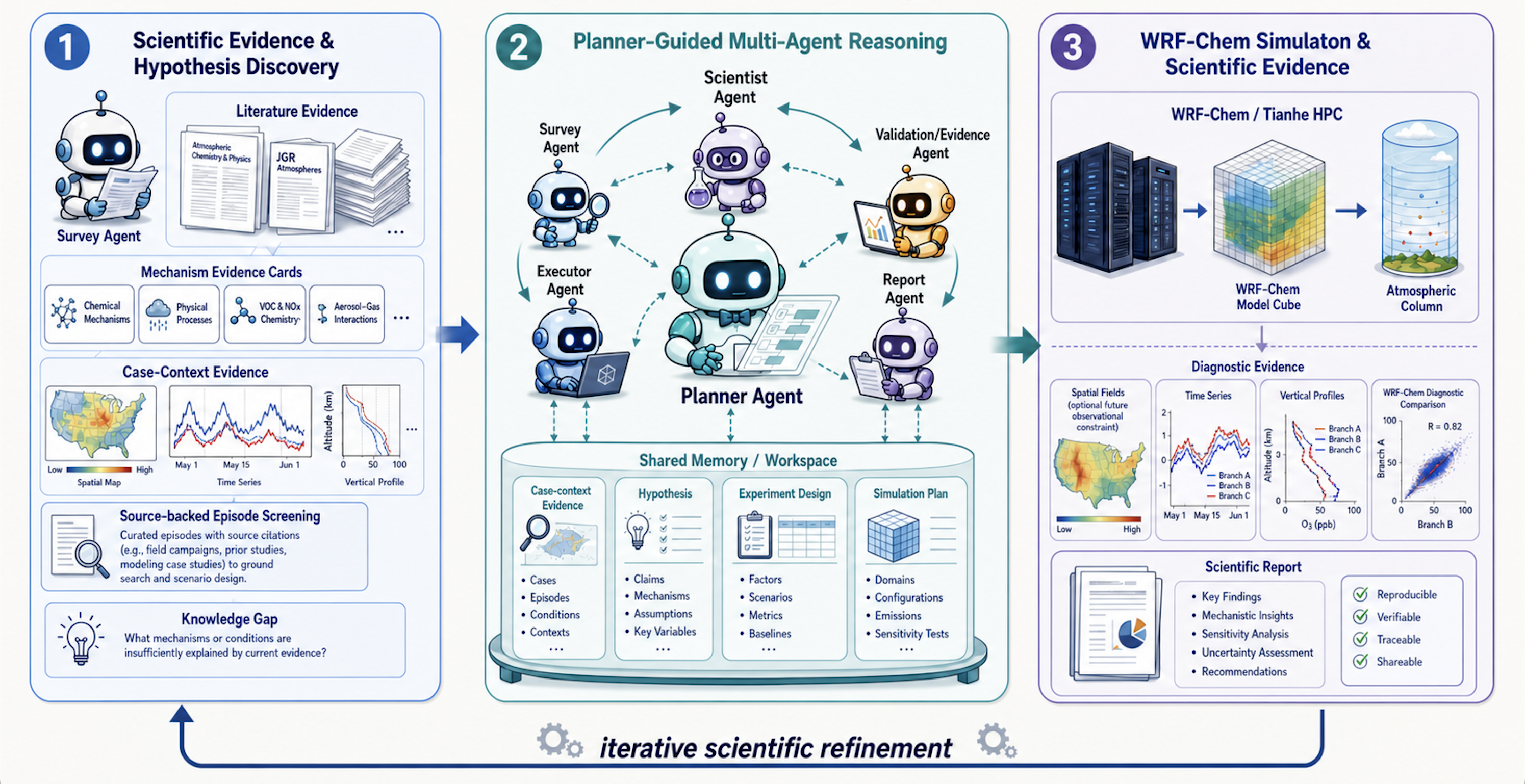}
  \caption{\textbf{End-to-end multi-agent architecture of \system.} A coordination layer organizes an open-ended atmospheric environmental problem across literature synthesis, hypothesis organization, \wrfchem{} experiment design, diagnostic evidence, and report expression. Diagnostic results, evidence gaps, and scientific evaluation can be fed back as targeted resurvey objectives for later hypothesis refinement and experiment design.}
  \label{fig:architecture}
\end{figure}

The multi-agent design is intended to represent distinct epistemic functions in mechanism-oriented research, rather than merely partitioning an implementation pipeline. Literature evidence synthesis, mechanistic hypothesis formation, experiment construction, model-execution organization, evidence diagnosis, and report expression are assigned to different research roles, while a coordination layer maintains the continuity of the inquiry. This division helps the system preserve the relation between a mechanism question and the evidence needed to test it (Fig.~\ref{fig:architecture}).

\subsection{Evidence-Constrained Hypothesis Generation}

The first capability of an autonomous AI scientist is not to generate generic research ideas. It is to turn an open atmospheric-environmental question into a literature-grounded and testable mechanism proposition. Starting from a pollutant, region, or candidate process, \system{} formulates targeted survey objectives around relevant mechanisms, applicable conditions, known uncertainties, and diagnostic variables. The Survey Agent then retrieves and synthesizes literature evidence, distinguishing support, conflict, and contextual constraints rather than treating papers as undifferentiated background material.

\begin{figure}[H]
  \centering
  \includegraphics[width=0.98\linewidth]{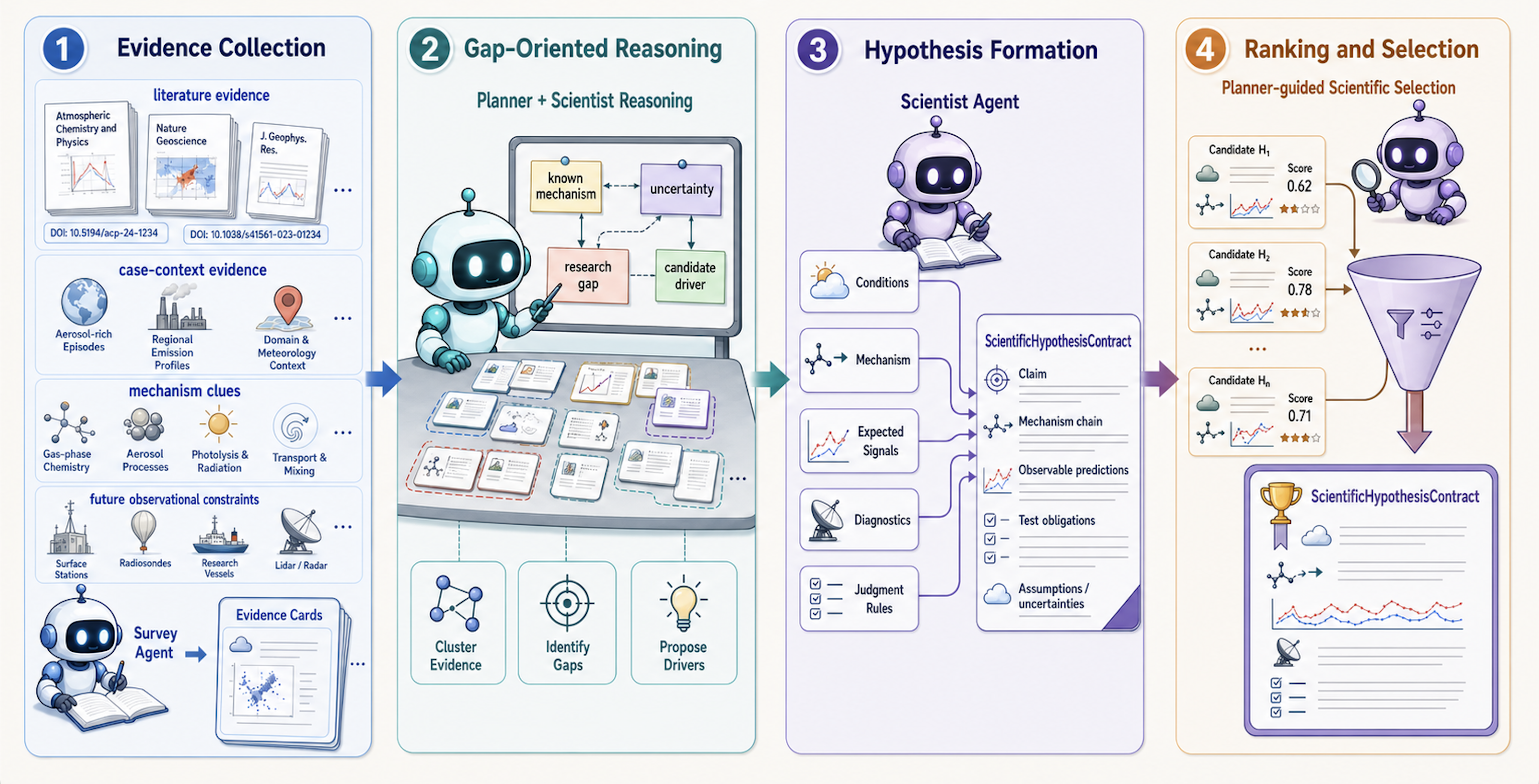}
  \caption{\textbf{Evidence-grounded formulation of testable hypotheses for atmospheric environmental mechanism research.} Open research questions are first translated into targeted literature-survey objectives around mechanisms, regional conditions, uncertainties, and diagnostic evidence. Literature evidence, case background, mechanistic clues, and future observational constraints are then organized into a traceable evidence basis. Candidate or already obtained mechanism propositions are expressed as structured hypotheses and passed to subsequent \wrfchem{} experiment design after assessment of testability, mechanism clarity, translatability to \wrfchem{} branch experiments, and evidence requirements.}
  \label{fig:hypothesis_discovery}
\end{figure}

This literature survey provides the basis for hypothesis formulation. The system organizes evidence according to relevance to the research question, mechanistic information content, and evidence strength, and identifies knowledge gaps, candidate drivers, expected diagnostic signals, and assumptions. The Scientist Agent uses this evidence base to form candidate hypotheses as causal chains, such as aerosol radiative perturbation, photolysis or boundary-layer change, and pollutant response. Each candidate is expressed with a mechanism claim, expected intermediate responses, possible falsification conditions, and observable or simulated diagnostics needed to support, weaken, or refute the mechanism (Fig.~\ref{fig:hypothesis_discovery}). When input information is incomplete, missing information is retained as uncertainty or a human-review point rather than being completed as fact.

For open-ended research directions, candidate hypotheses are iteratively assessed for testability, mechanism clarity, translatability to \wrfchem{} branch experiments, falsifiability, and evidence requirements. High-priority candidates are not passed directly to validation. They are refined by specifying missing causal links, adding competing mechanisms, and triggering targeted resurvey when the evidence basis is insufficient. The hypothesis selected for experiment design is therefore not simply the highest-scoring text, but a research proposition filtered jointly by evidence support, mechanistic discriminability, and numerical-experiment feasibility. These evidence weights act only as auxiliary signals; the degree of support for a hypothesis is determined by subsequent branch experiments, diagnostic evidence, and scientific evaluation.

\subsection{Experiments and Evidence}

\begin{figure}[H]
  \centering
  \includegraphics[width=0.98\linewidth]{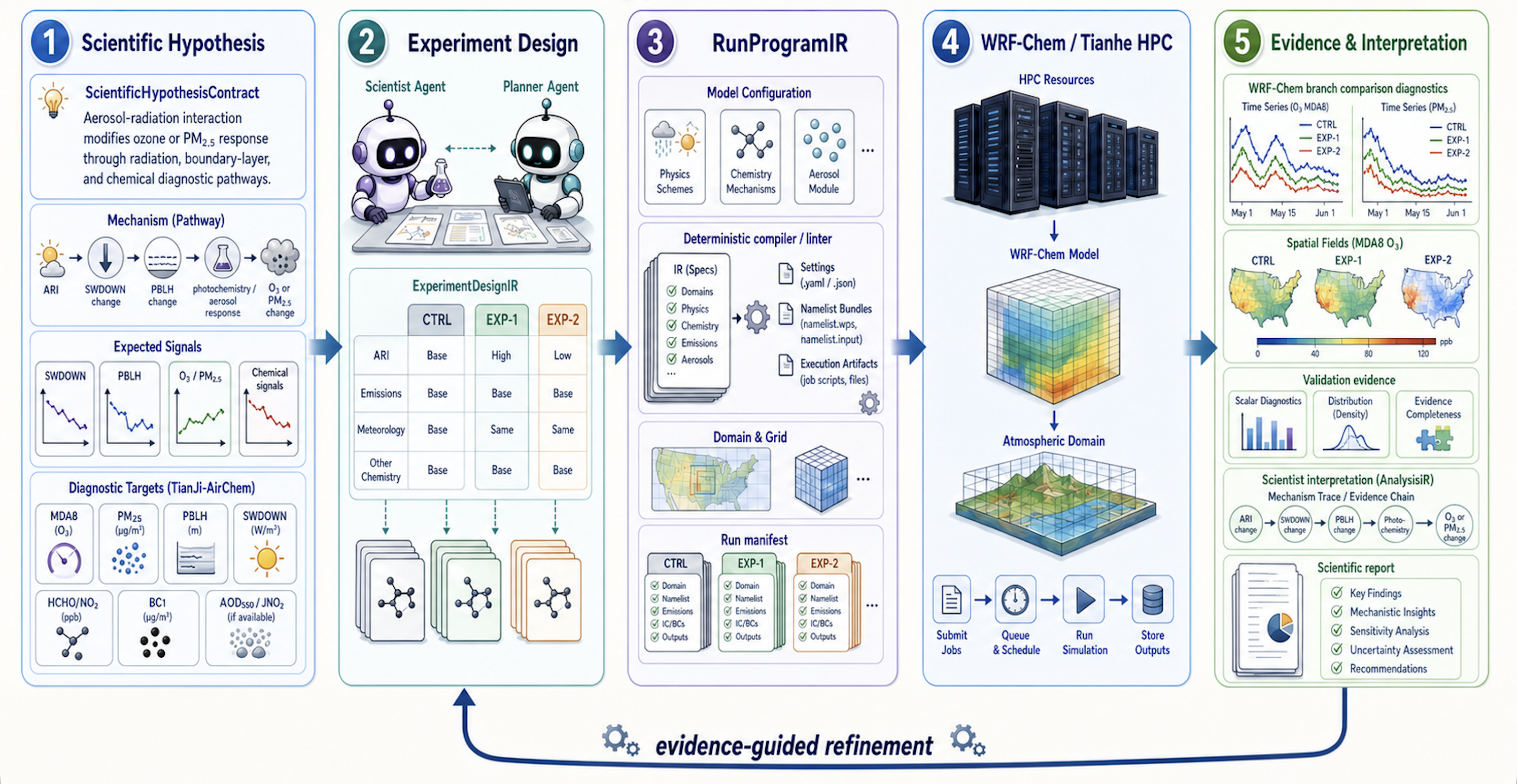}
  \caption{\textbf{Closed-loop chain from mechanistic hypothesis to \wrfchem{} evidence judgement.} The system decomposes a mechanism proposition into antecedent conditions, perturbable processes, branch contrasts, diagnostic variables, and evidence criteria. Model execution, diagnostic extraction, evidence gaps, and scientific judgement can therefore be traced back to the causal links specified by the hypothesis.}
  \label{fig:research_chain}
\end{figure}

After a hypothesis is obtained, \system{} expresses the mechanistic proposition as a set of controlled and comparable \wrfchem{} branch experiments. The design records control and sensitivity branches, perturbation factors, target pollutants, intermediate mechanism variables, and diagnostics required for judgement. For ozone problems, the relevant diagnostics commonly include \mda{}, \swdown{}, \pblh{}, photolysis rates, and chemical-sensitivity indicators. For \pmfine{} problems, they include particulate concentration, radiation variables, boundary-layer height, thermal structure, and aerosol components (Fig.~\ref{fig:research_chain}).

The translation begins by decomposing the hypothesis into an antecedent condition, a perturbable process, a proposed mechanism chain, a target pollutant response, and the evidence needed to judge each link. The antecedent condition determines the region, season, episode type, and baseline simulation. The perturbable process determines which sensitivity branch should be introduced, such as an emission scaling, aerosol--radiation interaction, or absorbing-aerosol perturbation. The mechanism chain determines the intermediate diagnostics: radiation or photolysis variables for radiative pathways, boundary-layer and thermal variables for mixing pathways, and concentration or composition variables for pollutant response. The target response defines the final branch contrast, while the evidence criteria specify which outcomes would support, weaken, or leave the hypothesis unresolved.

This mapping follows three principles. The first is contrast: each key mechanism claim should correspond to at least one branch difference that isolates the relevant process. The second is diagnosability: the experiment should examine not only the target pollutant concentration but also intermediate variables in the proposed mechanism chain. The third is falsifiability: the design records which outcomes would weaken or refute the hypothesis before the run is executed. Maintaining consistency across branches is scientifically important because it allows subsequent interpretation to focus on the intended process perturbation rather than incidental differences in experimental setup.

At the numerical-experiment stage, the system turns what needs to be tested scientifically into a set of runnable and comparable \wrfchem{} branches, so that branch differences mainly arise from specified process perturbations. In this sense, \wrfchem{} is not used merely to reproduce a pollution episode, but as a mechanism-testing platform. Scientist and Planner roles define the research intent, including region, period, perturbation factors, branch relations, and target diagnostics. Preprocessing and execution roles then examine whether meteorological inputs, emissions processing, chemistry and physics options, runtime resources, and branch consistency are adequate for the intended contrast.

Evidence evaluation is likewise not a simple threshold rule. The analysis first asks whether the intended perturbation is effectively expressed in model results. It then assesses whether intermediate radiative, boundary-layer, thermal, or chemical variables change along the direction expected by the hypothesis, before evaluating the target pollutant response. Model outputs are therefore summarized as branch differences, effect directions, effect magnitudes, and evidence gaps. When the target response is weak, key diagnostics are missing, or a core perturbation is not expressed, the outcome is summarized as partial support, rejection, or insufficient evidence, together with the diagnostics and experiment directions needed in the next round. This feedback is particularly important because a non-supportive result often does not imply that the physical mechanism is impossible; it may indicate that emissions perturbations, vertical structure, optical properties, photolysis diagnostics, or observational constraints need to be better specified.

\section{Case Studies and Validation}

\subsection{Case I: Aerosol--Radiation Interaction and Ozone Response}

The H1 case tests uncertainty in ozone response to a \nox{} emission-reduction perturbation under high-AOD stagnant summertime conditions over the North China Plain. Near-surface ozone responds nonlinearly to \nox{} reductions. Under \nox-limited conditions, reducing \nox{} usually suppresses ozone production. Under \voc-limited or \nox-saturated conditions, it may weaken titration or alter radical cycling. Ozone decreases can therefore be weak, or local increases may occur \cite{sillman1995,sillman1999,duncan2010omi,li2022indicators,jin2015hcho}. This study does not assume that the difference between \nox-cut and CTRL represents a net ozone-reduction benefit. It treats that difference as the ozone-response signal under a specified \nox{} reduction magnitude. Under high-AOD stagnant conditions, aerosol--radiation interaction (ARI) may weaken shortwave radiation, reduce photolysis rates, and suppress boundary-layer development \cite{yang2022apiarf,gao2022twoway}. These changes can reshape the photochemical and mixing environment for ozone production. We express this mechanism as H1: ARI may modulate the \mda{} response under a specific \nox{} reduction magnitude by changing radiation, photolysis, and boundary-layer conditions.

\begin{figure}[H]
  \centering
  \includegraphics[width=0.98\linewidth]{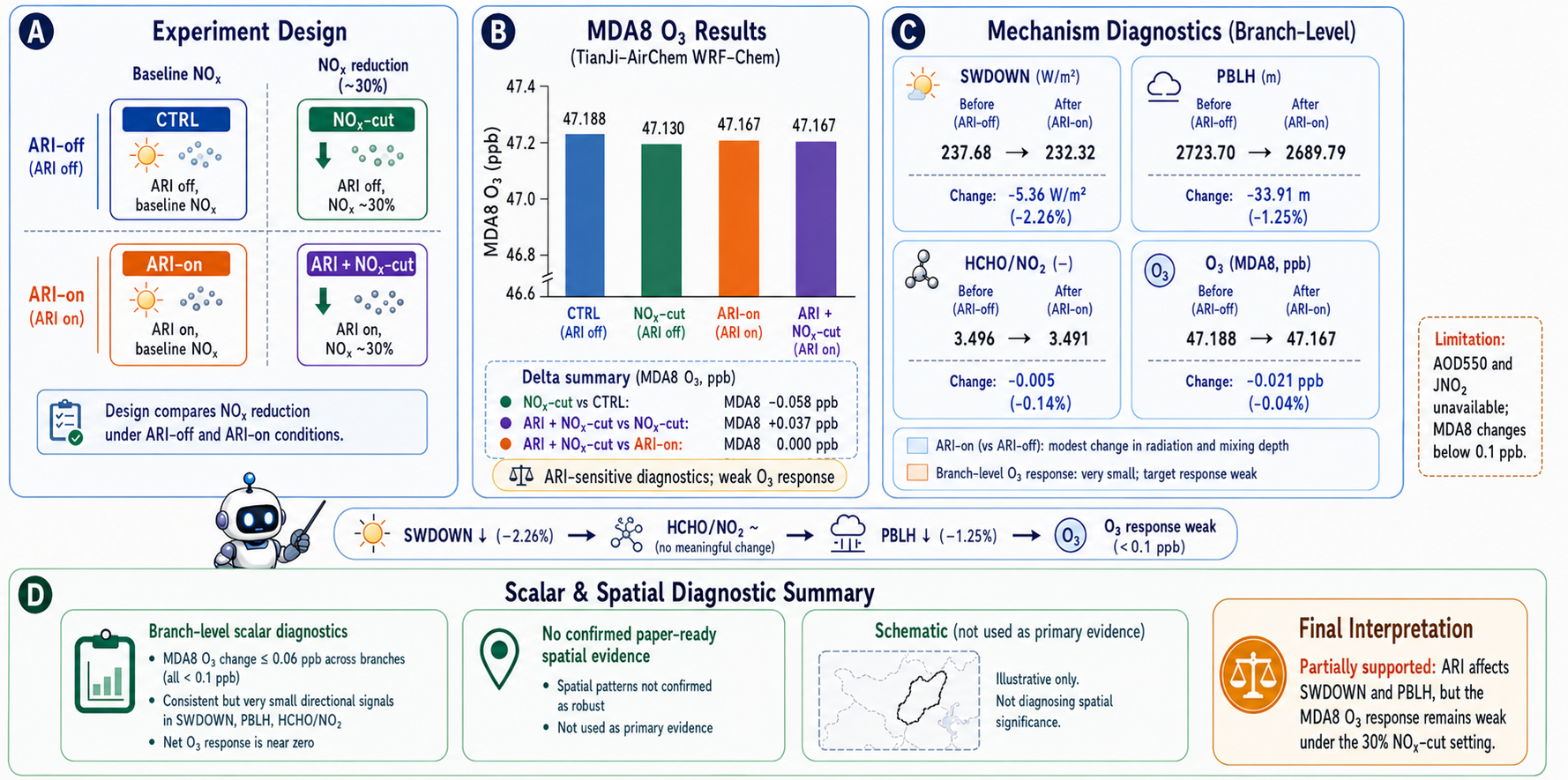}
  \caption{\textbf{Integrated diagnosis of the H1 branch experiment for summertime ozone response to \nox{} reduction over the North China Plain.} The four branches separate the effects of an approximately 30\% \nox{} reduction, ARI activation, and their combination. ARI effects on \swdown{} and \pblh{} are directionally consistent with the expected aerosol radiative feedback that weakens photochemical and boundary-layer mixing conditions, but the \mda{} response is smaller than 0.1 ppb, and key optical and photolysis diagnostics such as AOD550 and JNO$_2$ are missing.}
  \label{fig:h1_result}
\end{figure}

Testing H1 requires separation of three effects. These are the direct ozone response to an approximately 30\% \nox{} reduction, ARI-induced changes in radiation and boundary-layer conditions, and any ARI modulation of the \nox{}-reduction response. \system{} therefore organized four \wrfchem{} branches: CTRL, \nox-cut, ARI-on, and ARI+\nox-cut. In \nox-cut and ARI+\nox-cut, the \nox{} emission scaling factor was 0.7. CTRL and ARI-on retained baseline \nox{} emissions. This reduction magnitude is essential for interpreting the branch differences. The results are not extrapolated to all possible \nox{} reduction strengths. The four branches allow emission perturbation and radiative feedback to be compared within the same experimental framework (Fig.~\ref{fig:h1_result}). The simulation used a single domain over the North China Plain, centred at approximately 32.0$^\circ$N and 118.0$^\circ$E. The grid was 100$\times$100, with 27 km horizontal resolution and 35 vertical layers. The simulation period was 15--18 June 2021, and the analysis period was 16--18 June 2021.

\begin{figure}[H]
  \centering
  \includegraphics[width=0.98\linewidth]{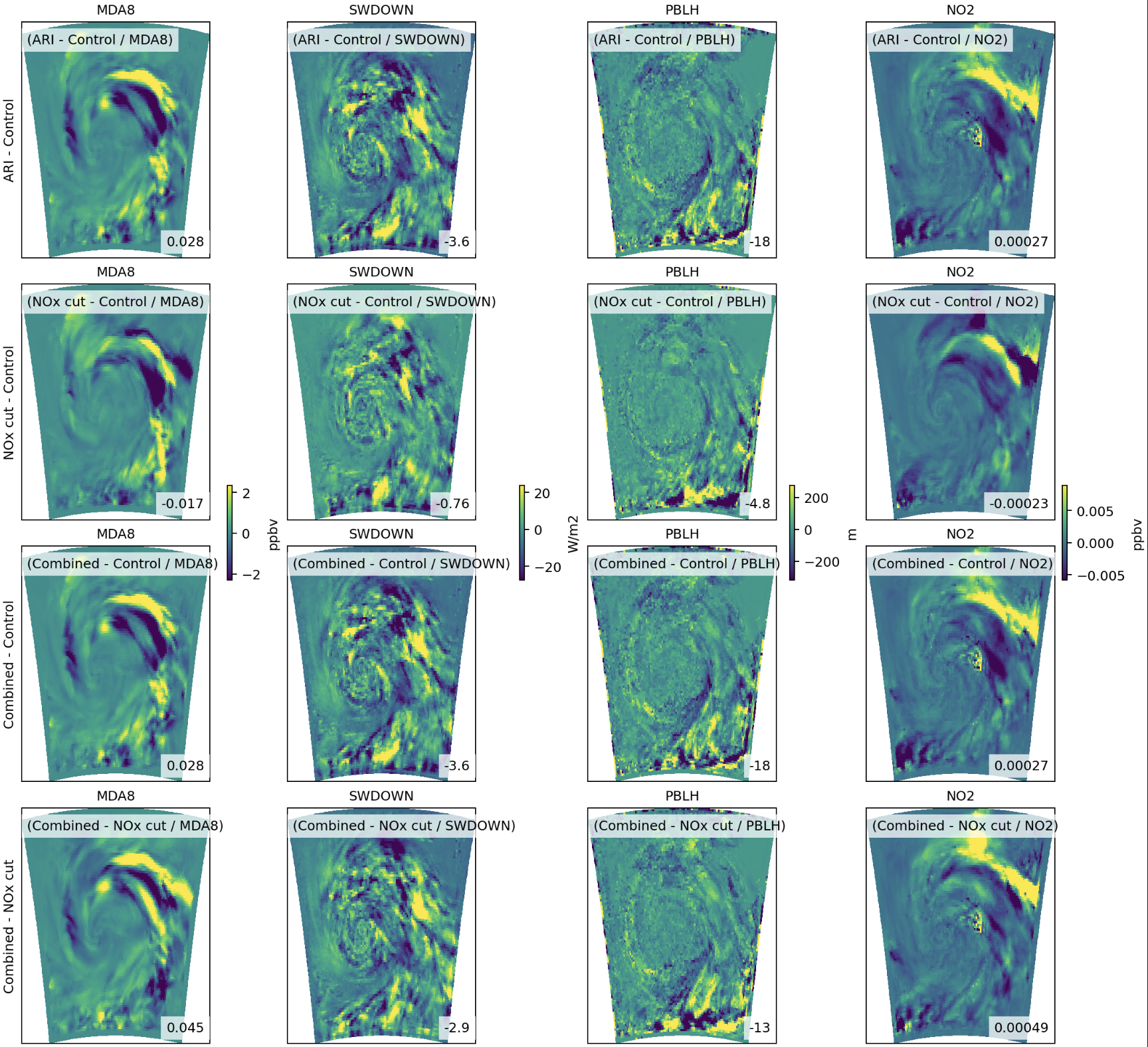}
  \caption{\textbf{Spatial-response diagnosis for the H1 branch experiment.} The figure compares spatial differences in \mda, \swdown, \pblh, and NO$_2$ for ARI, \nox{} reduction, and combined perturbations relative to the control branch, and also shows the combined branch relative to the \nox{}-cut branch. \swdown{} and \pblh{} show clear but spatially heterogeneous perturbation structures, whereas the \mda{} response is weak and spatially complex.}
  \label{fig:h1_spatial}
\end{figure}

The central H1 result is a separation between intermediate mechanism variables and the target ozone response. Relative to CTRL, ARI-on decreased \swdown{} and \pblh{} by approximately 2.26\% and 1.25\%, respectively. These directions are consistent with aerosol radiative feedback weakening photochemistry and boundary-layer mixing. However, this physical-environment change did not translate into a clear negative ozone response. Relative to CTRL, \nox-cut reduced \mda{} by only about 0.058 ppb. ARI-on changed \mda{} by about $-0.021$ ppb, and ARI+\nox-cut was almost identical to ARI-on. Under the 30\% \nox{} reduction setting, the ozone response was therefore very small. It should not be interpreted as evidence for a robust ozone-decreasing effect of \nox{} reduction in this case. Spatial diagnostics showed heterogeneous structures in \mda{}, \swdown{}, \pblh{}, and NO$_2$ for the ARI, \nox{}-reduction, and combined perturbations. The ozone response remained weak and spatially complex (Fig.~\ref{fig:h1_spatial}).

This case illustrates the bounded nature of the evidence interpretation. Intermediate mechanism variables provided directional support, but the target-pollutant response was weak and key diagnostics were missing. H1 can therefore only be summarized as partially supported. ARI effects on radiation and boundary-layer conditions are identifiable in the current results, but its modulation of the response to a 30\% \nox{} reduction lacks strong evidence. Subsequent work should add AOD550 and JNO$_2$ diagnostics, test multiple \nox{} reduction levels, and screen more representative pollution episodes.

\subsection{Case II: Black Carbon Feedback and Wintertime PM$_{2.5}$}

The H2 case examines the possible role of black-carbon (BC) absorbing aerosol feedback under stagnant wintertime heating conditions in the Guanzhong Basin. Winter pollution episodes in this region often include high emissions, weak winds, a shallow boundary layer, and stable stratification. The radiative effect of absorbing BC does not simply alter near-surface pollutant concentration. Its vertical distribution can reshape the atmospheric heating structure. Reduced shortwave radiation tends to cool the near surface. An elevated BC-containing aerosol layer can absorb radiation and warm the air aloft. Together, these effects may strengthen vertical stability, depress the boundary layer, suppress turbulent diffusion, and form a pollution-trapping dome effect \cite{petaja2016feedback,ding2016blackcarbon,wang2018dome}. We express this mechanism as H2: if the BC concentration magnitude and vertical distribution form an effective absorbing heating layer, BC absorbing ARI may amplify the \pmfine{} pollution response.

\begin{figure}[H]
  \centering
  \includegraphics[width=0.98\linewidth]{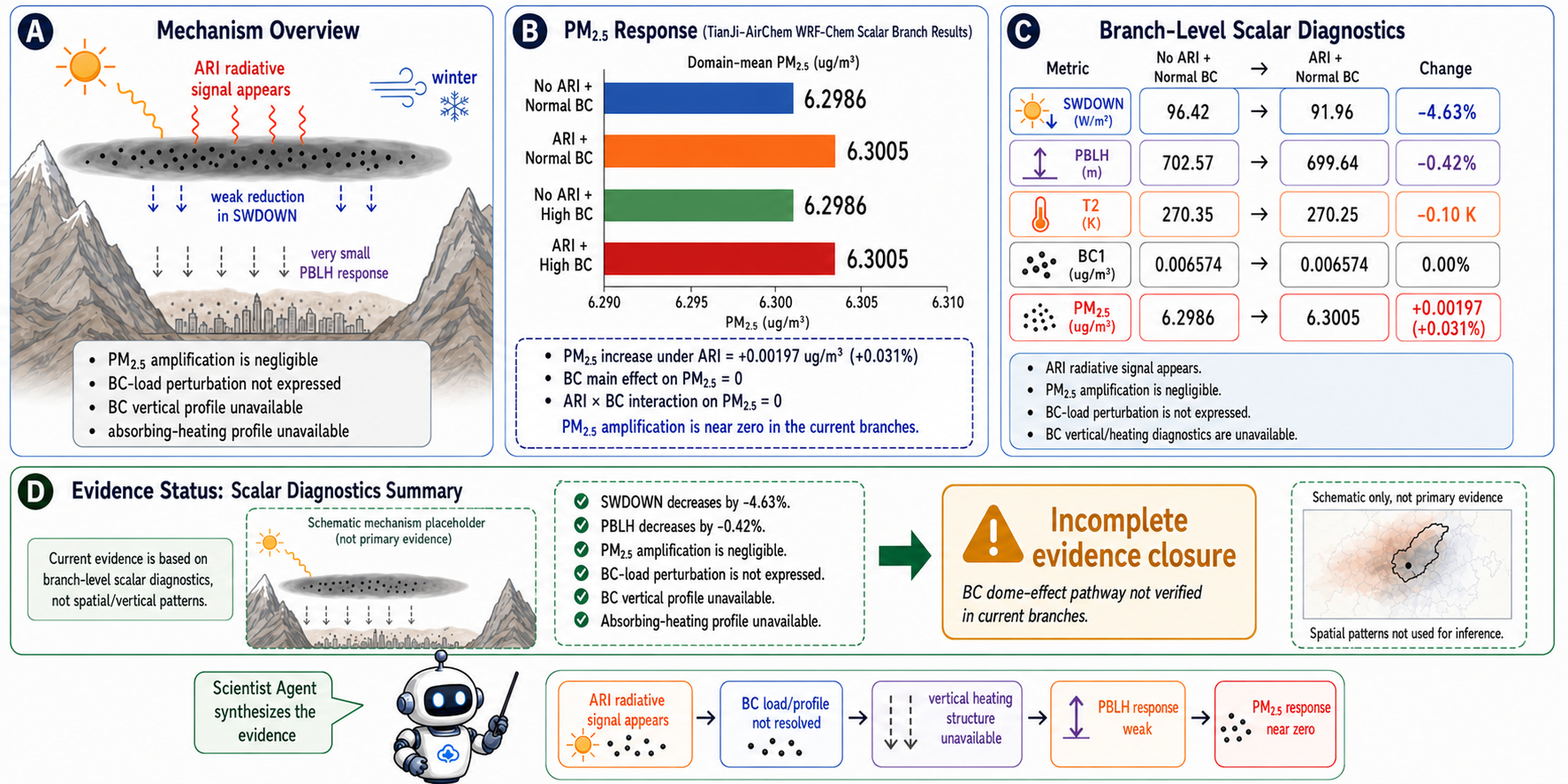}
  \caption{\textbf{Integrated diagnosis of the H2 branch experiment for the wintertime black-carbon absorbing-feedback hypothesis in the Guanzhong Basin.} The 2$\times$2 factorial branches compare the effects of ARI on/off and BC-load perturbation on \swdown, \pblh, BC1, and \pmfine. The ARI branch produces a shortwave-radiation reduction signal, but the \pmfine{} response is close to zero. The high-BC branch is almost identical to the normal-BC branch, indicating that the current results do not effectively express the BC-load perturbation or its vertical absorbing-heating structure.}
  \label{fig:h2_result}
\end{figure}

The key test is whether a radiative perturbation propagates into vertical thermal structure, boundary-layer stability, turbulent diffusion, and near-surface particulate accumulation. Demonstrating this mechanism therefore requires more than \swdown{}, \pblh{}, and \pmfine{} comparisons. It also requires evidence for BC concentration magnitude, vertical distribution, absorbing-heating profiles, and vertical temperature response. To test this transmission chain, \system{} adopted a 2$\times$2 factorial branch design. The branches combined ARI on/off with BC-load perturbation: No-ARI/normal-BC, ARI/normal-BC, No-ARI/high-BC, and ARI/high-BC (Fig.~\ref{fig:h2_result}). The simulation used a single domain over the Guanzhong Basin, centred at approximately 34.2$^\circ$N and 108.8$^\circ$E. The grid was 99$\times$99, with 27 km horizontal resolution and 35 vertical layers. The simulation period was 10--12 December 2014.

\begin{figure}[H]
  \centering
  \includegraphics[width=0.98\linewidth]{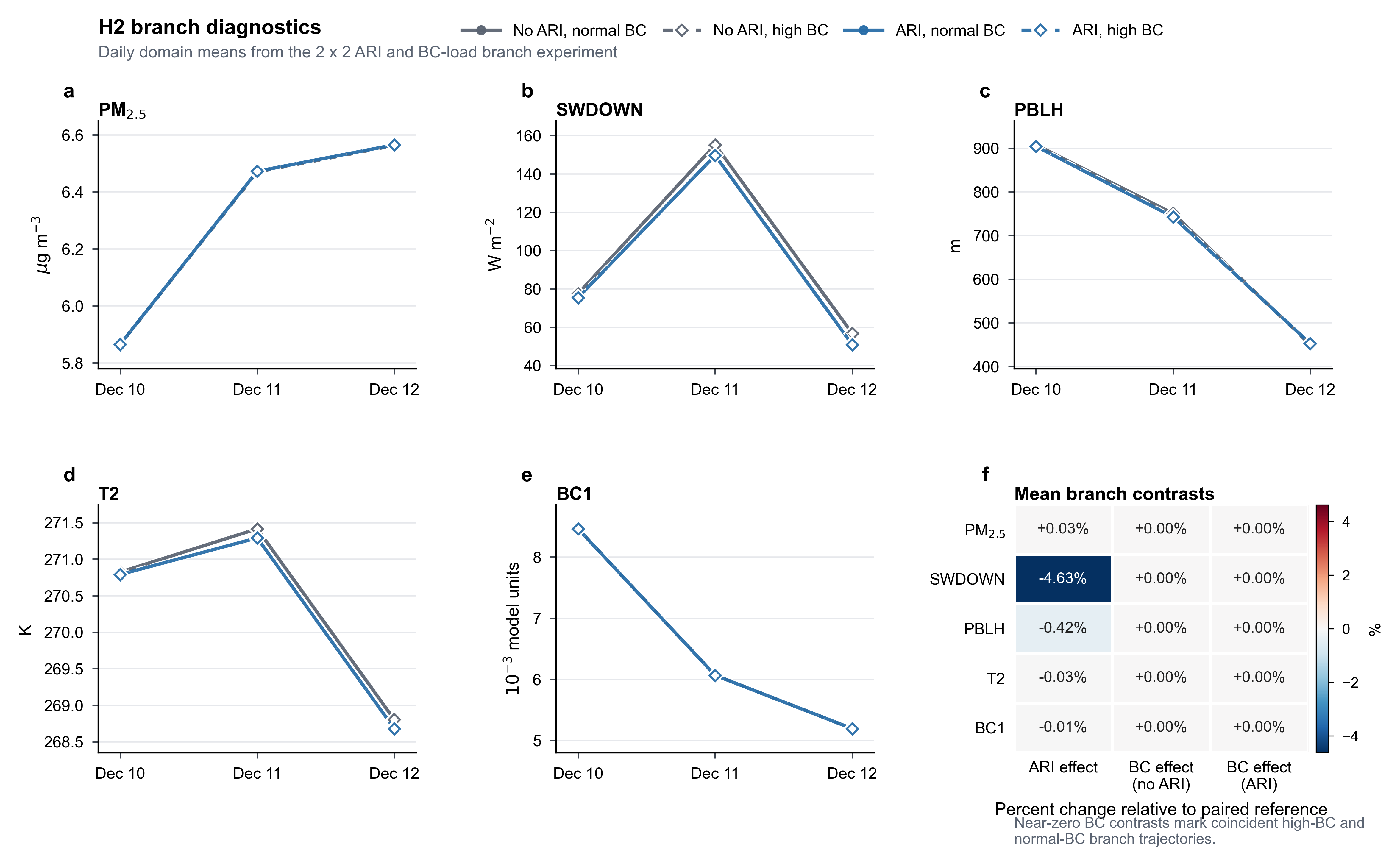}
  \caption{\textbf{Daily branch time-series diagnostics for the H2 experiment.} \textbf{a--e}, Domain-mean daily trajectories of \pmfine, \swdown, \pblh, T2, and BC1 across the four ARI $\times$ BC-load branches. Grey denotes no-ARI branches and blue denotes ARI branches; solid circles denote normal-BC branches and open diamonds denote high-BC branches. Coincident normal-BC and high-BC trajectories indicate that the current high-BC perturbation is not expressed in these diagnostics. \textbf{f}, Mean branch-contrast matrix expressed as percent change relative to the paired reference branch. The ARI contrast shows the main radiative response, whereas the two BC contrasts remain near zero.}
  \label{fig:h2_timeseries_diagnostics}
\end{figure}

The current H2 experiment did not capture a complete evidence chain for the BC dome effect. The ARI branch produced a recognizable radiative signal. \swdown{} decreased by about 4.63\%, and \pblh{} decreased slightly by about 0.42\%. However, the \pmfine{} increase was only about 0.00197 $\mu$g m$^{-3}$, close to zero. The high-BC branch was also almost identical to the normal-BC branch. This indicates that the current BC-load perturbation was not effectively expressed in the model results. Because differences in BC concentration magnitude, vertical distribution, and absorbing-heating profiles were missing, the current results cannot demonstrate the full dome-effect chain. The daily diagnostic trajectories and mean-contrast matrix make this evidence break visible across the target and mechanism variables (Fig.~\ref{fig:h2_timeseries_diagnostics}).

The main scientific information from this experiment is not a denial of the physical possibility of a BC-induced absorbing-aerosol dome effect. Rather, it shows that the current branch results do not express the BC-load perturbation, vertical distribution, or absorbing-heating profile needed to test that mechanism. The evidence gaps are concentrated in three links: ineffective expression of the BC-load perturbation, missing BC vertical-distribution and absorbing-heating-profile evidence, and a near-zero \pmfine{} response. Subsequent work should therefore check whether the BC emission perturbation enters the model, whether the BC vertical profile is reasonable, whether absorption heating warms the air aloft, and whether boundary-layer depression follows.

\subsection{Mechanistic Evidence Assessment}

H1 and H2 together show that \system{} treats hypothesis validation as an evidence-state problem rather than a binary decision. H1 represents directional mechanism evidence with insufficient closure: radiation and boundary-layer variables respond as expected, but the ozone signal is weak and optical and photolysis diagnostics are missing. H2 represents a different state, in which the current experiment does not express the core BC perturbation and vertical absorbing structure required to test the proposed mechanism. In human-led research, such distinctions are often scattered across run notes, scripts, and figures. Here they become comparable scientific judgements that can guide subsequent diagnostics and branch design without claiming that a single non-supportive run disproves the physical mechanism itself.

\section{System-Level Analysis}

\subsection{Reliability of Research Actions}

\begin{figure}[H]
  \centering
  \includegraphics[width=0.98\linewidth]{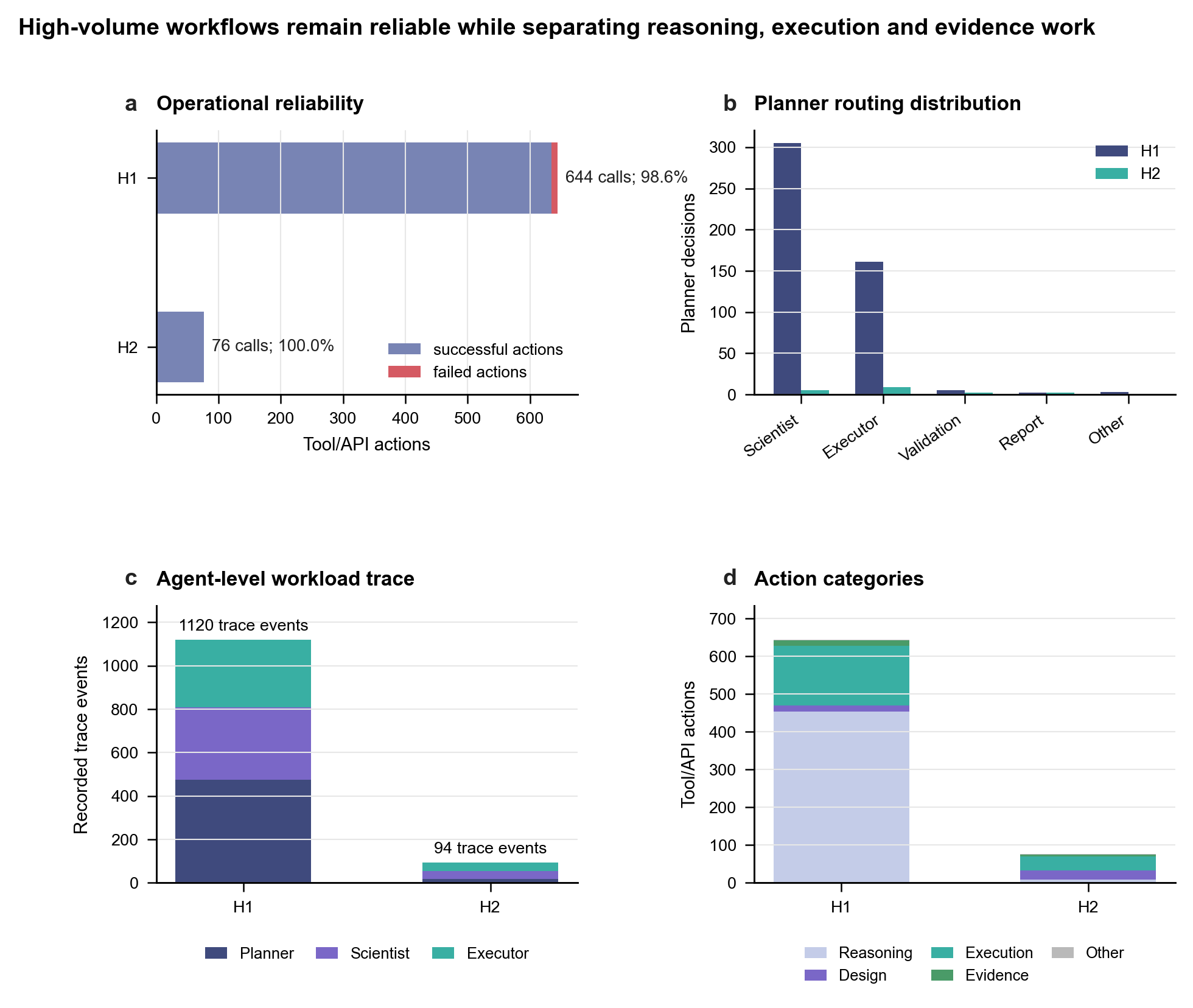}
  \caption{\textbf{Research-action reliability and multi-agent workload during the H1 and H2 runs.} \textbf{a}, Scale and success rate of tool/API-mediated research actions in the two cases. \textbf{b}, Distribution of Planner routing decisions across target roles. \textbf{c}, Agent-level event counts recorded in the process trace. \textbf{d}, Distribution of tool/API actions across reasoning, design, execution, evidence, and other categories.}
  \label{fig:workloadreliability}
\end{figure}

For an autonomous AI scientist, reliability depends on more than final textual conclusions. Long-chain research actions must remain stable, recoverable, and inspectable, especially when the workflow connects hypothesis organization, numerical experimentation, evidence analysis, and interpretation. The H1 process record contains 644 tool-mediated research actions, of which 635 succeeded and 9 failed, giving a success rate of 98.6\%. The H2 process record contains 76 such actions, all successful (Fig.~\ref{fig:workloadreliability}a). These records provide process-level evidence that the system maintained continuity across experiment design, remote execution, evidence analysis, and report expression (Fig.~\ref{fig:workloadreliability}b--d).

\subsection{Cross-Stage Coordination}

\begin{figure}[H]
  \centering
  \includegraphics[width=0.98\linewidth]{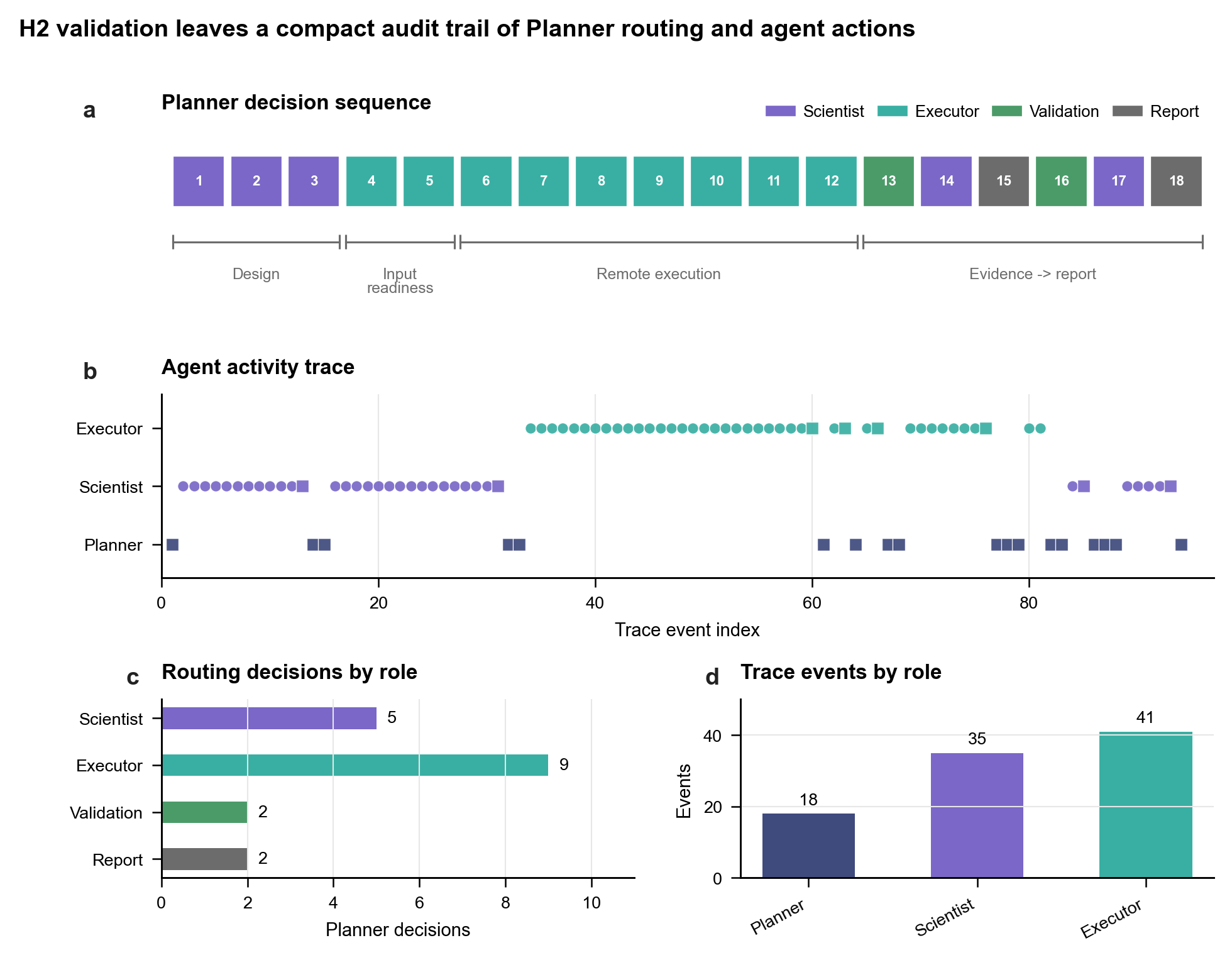}
  \caption{\textbf{Coordination trajectory across research stages during the H2 run.} \textbf{a}, Sequence of 18 Planner routing decisions from experiment design and input readiness to remote execution and evidence-to-report synthesis. \textbf{b}, Planner, Scientist, and Executor activity across trace-event indices. \textbf{c}, Number of routing decisions directed to each target role. \textbf{d}, Event counts by role in the process trace.}
  \label{fig:routing_handoff}
\end{figure}

The H2 trajectory further shows how a complex mechanism-validation task is coordinated across research stages. The overall distribution indicates that most \wrfchem{} validation effort is concentrated in experiment design and model-execution organization (Fig.~\ref{fig:workloadreliability}b). Within a single H2 run, research actions moved from experiment design and input readiness to remote execution, evidence synthesis, scientific evaluation, and report generation. The corresponding trace shows a clear collaboration path from early-stage design to late-stage interpretation (Fig.~\ref{fig:routing_handoff}). These figures describe the research behaviour of the AI scientist rather than pollutant outcomes themselves.

\subsection{Diagnostic Tasks}

\begin{figure}[H]
  \centering
  \includegraphics[width=0.98\linewidth]{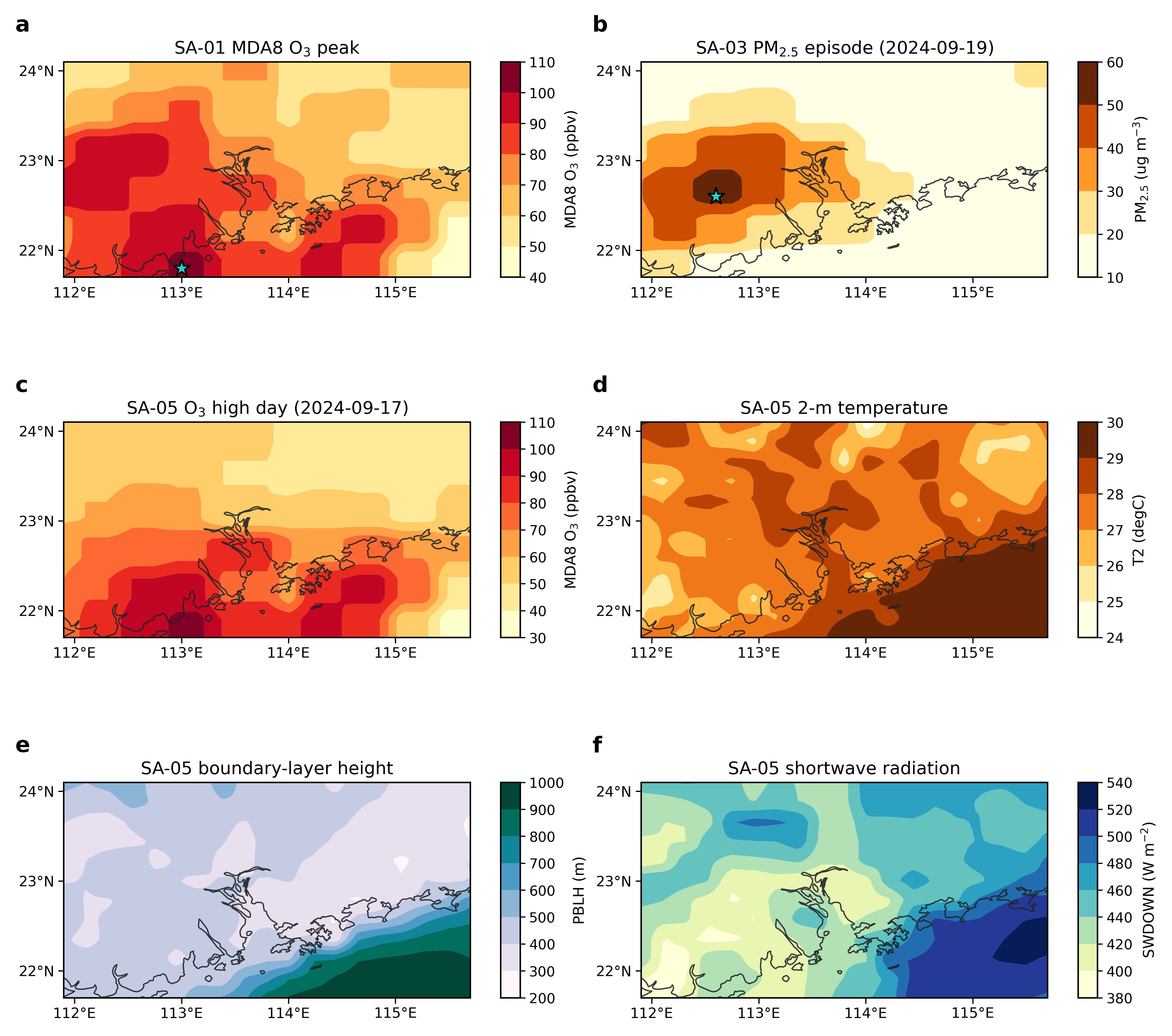}
  \caption{\textbf{Examples from the diagnostic tasks.} \textbf{a}, SA-01 \mda{} peak distribution and peak-location identification. \textbf{b}, SA-03 \pmfine{} episode diagnosis, showing regional episode selection and pollution-centre identification. \textbf{c--f}, SA-05 O$_3$--meteorology co-variation diagnosis, showing high-ozone-day \mda{}, T2, \pblh{}, and \swdown{} patterns.}
  \label{fig:diagnostic_tasks}
\end{figure}

The diagnostic tasks evaluate whether lightweight data analysis can be expressed as qualified scientific interpretation rather than isolated plotting operations. The three tasks cover \mda{} peak identification, \pmfine{} episode spatial diagnosis, and O$_3$--meteorology co-variation analysis. SA-01 calculated rolling 8-h O$_3$ and extracted \mda{}, followed by peak-window and peak-location identification. The system identified a peak \mda{} value of 102.7 ppbv, with the peak window ending at 2024-09-17 11:00 UTC and the peak location near 21.8$^\circ$N, 113.0$^\circ$E. SA-03 selected a regional \pmfine{} episode from hourly data and mapped its spatial footprint. The system selected 2024-09-19 as the domain-mean \pmfine{} episode, with a domain mean of 22.11 $\mu$g m$^{-3}$ and a peak-location daily mean of 54.67 $\mu$g m$^{-3}$. SA-05 jointly inspected O$_3$, T2, \pblh{}, and \swdown{} on a high-ozone day. The system selected 2024-09-17 and reported spatial correlations between O$_3$ and T2, \pblh{}, and \swdown{} of 0.178, 0.209, and $-0.225$, respectively. These tasks show that \system{} can organize metric extraction, peak identification, episode selection, and co-variation patterns into constrained post-processing interpretations (Table~\ref{tab:diagnostic_tasks}; Fig.~\ref{fig:diagnostic_tasks}).

\begin{table}[H]
\centering
\caption{\textbf{Three representative post-processing and diagnostic tasks.}}
\label{tab:diagnostic_tasks}
\footnotesize
\setlength{\tabcolsep}{3pt}
\renewcommand{\arraystretch}{1.14}
\begin{tabular}{L{3.0cm} L{6.2cm} L{5.0cm}}
\toprule
\textbf{Diagnostic task} & \textbf{Analysis step} & \textbf{Scientific diagnostic capability} \\
\midrule
SA-01 \mda{} spatial peak
& Rolling 8-h mean; \mda{} extraction; peak-window and location search
& Metric extraction; temporal search; hotspot localisation \\
\midrule
SA-03 \pmfine{} episode spatial diagnosis
& Daily averaging of hourly \pmfine{} data; regional episode selection; peak-location search
& Pollution-event identification; spatial-footprint diagnosis; episode interpretation \\
\midrule
SA-05 O$_3$--meteorology co-variation diagnosis
& High-ozone-day selection; O$_3$, T2, \pblh{}, and \swdown{} spatial-pattern generation; spatial-correlation check
& Multi-variable co-variation diagnosis; qualified interpretation; meteorological-context synthesis \\
\bottomrule
\end{tabular}
\end{table}

\section{Discussion and Conclusion}

\subsection{TianJi-Environ as an AI Scientist}

The value of \system{} lies not in replacing \wrfchem{} or expert judgement, but in making the mechanism-validation process explicit, executable, and auditable. Atmospheric-chemistry mechanism studies often rely on expert coordination across hypotheses, model configurations, diagnostic variables, and interpretation. In such workflows, the reasoning that connects a mechanistic proposition to a qualified conclusion is frequently distributed across scripts, configuration files, diagnostic plots, and informal expert decisions. \system{} addresses this gap by organizing the hypothesis--experiment--diagnosis--evidence--conclusion process into a traceable research chain.

This distinguishes \system{} from prediction-oriented atmospheric AI systems. While data-driven prediction models primarily estimate future atmospheric states, \system{} targets a different scientific task: determining whether a hypothesized mechanism is supported, incomplete, or insufficiently constrained by model evidence. Given a mechanism hypothesis, the system organizes literature evidence, formulates testable process chains, generates \wrfchem{} experiments, evaluates diagnostic consistency, and localizes unsupported links. In this sense, \system{} acts as a domain-grounded AI Scientist for atmospheric-chemistry mechanism validation rather than a general-purpose agent workflow.

The two case studies illustrate this distinction. In the ozone-response case, the system identified physically consistent aerosol--radiation-interaction signals but did not over-interpret them as a closed ozone-control mechanism. In the \pmfine{} case, the system localized the evidence gap to the ineffective propagation of the black-carbon perturbation and the lack of vertical heating diagnostics. These results suggest that the main role of \system{} is not to produce affirmative conclusions, but to structure scientific uncertainty and make the evidential basis of mechanism interpretation reproducible.

\subsection{Limitations and Future Directions}

\system{} remains a prototype, and its current limitations mainly concern evidence constraints and iterative validation. First, the present implementation primarily constructs mechanistic evidence from controlled \wrfchem{} experiments. Independent observational constraints, including ground-based measurements, satellite AOD, reanalysis products, and vertical profile observations, have not yet been incorporated as a standard component of the validation workflow. Future versions should integrate model--observation consistency checks so that mechanism interpretation is constrained by both numerical experiments and external observational evidence.

Second, the current workflow mainly demonstrates a single-cycle path from hypothesis generation to mechanism validation. When the evidence chain remains incomplete, the system can identify unsupported links, but automated redesign of follow-up experiments is still limited. A more mature system should be able to propose additional diagnostics, revise perturbation designs, trigger new \wrfchem{} simulations, and update the evidence state through multiple validation cycles.

Future work will therefore focus on three directions. First, observational data, literature-derived constraints, and model diagnostics should be integrated into a unified evidence framework. Second, the system should be extended with active redesign capabilities, allowing it to respond to incomplete evidence chains by proposing targeted follow-up experiments. Third, the execution layer should further strengthen configuration constraints, runtime checks, and provenance tracking for complex \wrfchem{} experiments. These extensions would move \system{} from a single-cycle validation prototype toward a continuously improving AI research assistant for atmospheric chemistry.

\subsection{Conclusion}

This study presents \system{}, an auditable AI Scientist for atmospheric-chemistry mechanism validation. The framework connects evidence-constrained hypothesis generation, \wrfchem{}-based numerical experimentation, diagnostic assessment, and qualified scientific interpretation within a traceable multi-agent workflow. Through case studies on ozone response and particulate-matter feedback, \system{} demonstrates the ability to identify directionally consistent mechanism signals, detect incomplete evidence chains, and localize unsupported process links. Overall, \system{} provides a new paradigm for coupling multi-agent AI systems with complex atmospheric-chemistry models to construct reproducible mechanistic evidence chains.

\section*{Code and Test-Result Availability}

Curated test results, run-process records, and source code supporting the examples reported in this manuscript will be released and maintained in the \href{https://github.com/zwww-www/TianJi-Environ}{project GitHub repository}.

\end{document}